\documentclass[useAMS,usenatbib]{mn2e}
\usepackage{graphicx,times}
\usepackage{natbib}
\usepackage{amssymb,amsmath}
\bibpunct{(}{)}{;}{a}{}{,}
\newcommand{\beq}{\begin{equation}}
\newcommand{\eeq}{\end{equation}}
\newcommand{\bea}{\begin{array}}
\newcommand{\eea}{\end{array}}

\title[Terrestrial Planets Formation under Migration]{Terrestrial Planets Formation under Migration:
the Systems near 4:2:1 Mean Motion Resonance
\\}

\author[Z. Sun et al.]{Zhao Sun$^{1, 2}$, Jianghui Ji$^{1}$\thanks{E-mail:
jijh@pmo.ac.cn}, Su Wang$^{1}$, Sheng Jin$^{1}$ \\
$^{1}$CAS Key Laboratory of Planetary Sciences, Purple Mountain Observatory, Chinese Academy of Sciences, Nanjing 210008,
China\\
$^{2}$University of Chinese Academy of Sciences, Beijing 100049, China}

\begin{document}

\date{Accepted. Received; in original form}

\pagerange{\pageref{firstpage}--\pageref{lastpage}} \pubyear{2015}

\maketitle

\label{firstpage}

\begin{abstract}
In this work, we extensively investigate the formation of near 4:2:1 mean motion
resonances (MMRs) configuration by performing two sets of N-body simulations. We model
the eccentricity damping, gas drag, type I and type II planetary
migration of planetesimals, planetary embryos and giant planets in
the first sets.  For the simulations of
giant planets with type II migration, the massive terrestrial planets,
with a mass up to several Earth masses, are likely produced in the systems.
We further show that by shepherding and/or scattering mechanisms through Jovian
planet's type II migration, the terrestrial planets and giant
planets in the systems can be evolved into a near 4:2:1 MMRs. Moreover, the models are applicable to the
formation of Kepler-238 and 302 systems. In the second set, we study the 4:2:1 MMRs
formation in the terrestrial planetary systems, where the planets
undergo type I migration and eccentricity damping. By considering type I migration,
$\sim$ 17.1\% of the simulations indicate that terrestrial planets are evolved into 4:2:1 MMRs.
However, this probability should depend on the initial conditions of planets.
Hence, we conclude that both type I and type II migration can play a crucial role in close-in
terrestrial planet formation.

\end{abstract}


\section{Introduction}

The number of exoplanets has substantially increased in recent years.
At the time of writing, over 3500 exoplanets (see www.exoplanet.eu) are discovered,
mostly through radial velocity and transiting surveys.
In particular, as of Nov. 10, 2015, Kepler mission released approximately 1030 confirmed planets
, along with over 4696 transiting planetary candidates \citep{fres13}.
This may imply that the planets appear to be very common, orbiting
other stars beyond our own solar system.

The observations show that close-in (short-period) terrestrial
planets are typically common in the exoplanetary systems. Roughly
speaking, one third to half of solar-type (FGK) stars can host at
least one planet with a mass less than $10~M_\oplus$ and an orbital
period ranging from 50 to 100 days \citep{how10, may11}. The
frequency of short-period terrestrial planets is at least as high
around M stars as around FGK stars, or even higher \citep{how12,
bon13, fres13}. These terrestrial planets are usually found in
multi-planet systems on compact but non-resonant orbits
\citep{udry07, lov11, lis11}. However, from the statistical results
of Kepler release data, a great many of planet pairs in the systems are
observed to be in near-resonant configurations \citep{zhang10,
lee13, mar13, wang14, zhang14}, such as the 4:2:1 MMRs. For example,
Kepler-238 \citep{rowe14} and Kepler-302 \citep{rowe14} systems
harbor close-in giant planets in near mean motion resonance (MMR)
configuration with other planets.  Therefore, these exciting observations
motivate us to explore and understand the formation and evolution of the planetary systems,
especially to investigate the short-period terrestrial planet formation
and the configuration formation of the 4:2:1 MMRs.

Several models have been proposed to explain the formation of close-in
terrestrial planets \citep{raym08}, and the formation scenarios include in situ accretion,
orbital migration arising from planetary embryos (type I migration \citep{GT80})
and gas giant planets (type II migration \citep{lp86}), dynamical instabilities
in systems of multiple gas giant planets, tidal circularization of eccentric terrestrial planets,
and photo-evaporation of close-in giant planets.
For the systems that are composed of short-period terrestrial planets, their protoplanetary disks
are suggestive of very massive \citep{raym08, hansen12, hansen13, chiang13, raymond14}, thus the planets
can accrete in situ from a large number of planetesimals and planetary embryos in the disk.
The formation and final configuration of close-in terrestrial planets are closely related to the strength of type-I migration, and the suppression of type I migration is required in case of in-situ super-Earths formation \citep{Ogihara2009,Ogihara2015}.
The observed systems of hot super-Earths mostly can contain 20 - 40 $M_\oplus$ in mass
within a fraction of an AU of the host star \citep{batalha13}.
Star-planet tidal interactions may play a role in circularizing planets
in highly-eccentric orbits and therefore reduce their semi-major axis in the evolution \citep{ford06, fab07, beau12, dong13}.
However, \citet{raym08} suggested that tidal effect can produce hot super-Earths,
but only for those relatively massive planets ($\ga 5 M_\oplus$) with very small perihelion distances ($\la$ 0.025 AU),
and even then the inward movement in orbital distance is only
0.1-0.15 AU at most, therefore tides are not strong enough to move many of the Kepler planets to the
nowadays observed separations, and additional dissipative processes are at play \citep{lee13}.

Alternatively, terrestrial planet may form inside a migrating giant planet \citep{raym06}.
Gas-giants can shepherd planetesimals and embryos interior to their orbits as they migrate inward,
which can further collide and merge into Earth-like planets \citep{zhou05}.
\citet{man07} showed materials that have been shepherded interior to
the migrating giant planet by moving MMRs can accrete into close-in terrestrial planets.
In addition, \citet{izi14} indicated that fast-migrating super-Earths only weakly perturb
the planetesimals disk and planetary embryos, whereas slowly migrating super-Earths shepherd
rocky material interior to their orbits in resonances and push toward the star.
Moreover, the orbital migration and planet-planet scattering
play a vital role in producing short-period terrestrial planets \citep{brun05, terq07, raym08, cos14}.
According to core-accretion model, the planetary embryos in the terrestrial
planet formation region, and the solid cores of giant planets, are both formed within $\sim 1$\, Myr
from kilometer-sized planetesimals \citep{Saf69, wet80}. Subsequently,
the massive solid cores can further accrete gas from the protoplanetary disk to form gas-giants
\citep{Kok02, idalin04} at Myr timescale, before the disk disperses
\citep{hai01}. In the late stage of planet formation, after the gas disk clears, the giant
planets ceases migrating, the numerous planetesimals and planetary embryos in the disk
become turbulent due to dynamical stirring by gas-giants over hundreds of Myrs or
even longer. Consequently, frequent orbital crossings and giant impacts
are likely to occur, which may eventually yield short-period Earth-like planets \citep{cham01, raym04, zha09, ji11}.

\citet{raym06} investigated terrestrial planets formation under type
II migration and in the model they included a giant planet and gas
drag. Based on Raymond et al's model, \citet{man07} further
considered an additional non-migrating giant planet. In our earlier
study, we have investigated the 3:2 and 2:1 MMRs configuration
formation \citep{wang14} in the system observed by Kepler mission,
and simply considered the planets with their nominal masses that are
over a few Earth masses without any growth \citep{wjz12, wang14}. In
the present work, we aim to explore terrestrial planet formation
especially in the system with 4:2:1 configuration under migration by
performing N-body simulations. In short, we have carried out two
kinds of simulations, where in the first model we include the
presence of giant planet and as a comparison, in the second model we
consider the configuration formation in the system with only
terrestrial planets.

The paper is structured as follows. In Section 2, we describe
the adopted models of planetary formation, including the disk model and the planetary migration scenarios.
In Section 3 we present numerical setup and major results of our investigation.
Finally, we summarize the outcomes and give a brief discussion in Section 4.

\section{Models}
Following the empirical minimum-mass solar nebula (MMSN, Hayashi
1981) model, the surface density of solid disk at stellar distance
$a$ is described as

\begin{equation}
\Sigma_{\rm d}=10f_{\rm d}\gamma_{\rm ice}\left(\frac{a}{1{\rm
AU}}\right)^{-3/2} ~~{\rm g cm^{-2}},
\label{solid}
\end{equation}
where $f_d$ and $\gamma_{\rm ice}$ are the solid and the volatile
enhancement factor, respectively. $\gamma_{\rm ice}$ is 4.2 exterior
to the snow line or 1 interior to snow line. The profile of gas
density is given as

\begin{equation}
\Sigma_{\rm g}=2.4\times 10^3f_{\rm g}f_{\rm dep}\left(\frac{a}{1\rm
AU}\right)^{-3/2}~~{\rm g cm^{-2}}, \label{dens}
\end{equation}
where $f_{\rm g}$ and $f_{\rm dep}$ are the gas enhancement factor
and gas depletion factor, respectively. $f_{\rm
dep}$=$\exp(-t/\tau_{\rm dep})$, where $t$ is the time and the
timescale $\tau_{\rm dep}$ is about few million years \citep{hai01}.
Herein we adopt $\tau_{\rm dep}=10^6 ~ \rm yr$. The inner edge
of the gas disk locates at 0.1 AU.

As they are considered small bodies in our simulations, the planetesimals
will be affected by the aerodynamical drag of the gas around them
\citep{Ada76,Tan99}. The force that a planetesimal with a mass $m$
will suffer from aerodynamical drag is written as

\begin{equation} \textbf{F}_{\rm aero}=-\frac{1}{2}C_{\rm
D}\pi S^2\rho_{\rm g}|\textbf{U}|\textbf{U},
\end{equation}
where $\textbf{U}=\textbf{V}_{\rm k}-\textbf{V}_{\rm g}$ is the
relative velocity between the planetesimal's Keplerian motion
$\textbf{V}_{\rm k}$ and the gas motion $\textbf{V}_{\rm g}$.
$C_{\rm D}$ is the drag coefficient. If an object has a large
Reynold number, $C_{\rm D}$ can be taken as 0.5. $S$ and $\rho_{\rm
g}$ are the planetesimal's radius and gas density, respectively.

For a planetary embryo embedding in the gaseous disk, their mutual interactions
will lead to an eccentricity damping of the embryo with a timescale
$\tau_{\rm damp}$ written as \citep{Cre06}

\begin{eqnarray}
\tau_{\rm damp1}=\left(\frac{e}{\dot
e}\right)=\frac{Q_e}{0.78}\left(\frac{M_*}{m}\right)
\left(\frac{M_*}{\Sigma_g a^2}\right)\left(\frac{h}{r}\right)^4\Omega^{-1}
\nonumber\\
\times\left[1+\frac{1}{4}\left(e\frac{r}{h}\right)^3\right]
,~~~~~~~~~~~~~~~~~ \label{damp}
\end{eqnarray}
where $h,~r,~\Omega$, and $e$ are disk scale
height, distance from central star, Keplerian angular velocity, and the
eccentricity of the embryo, respectively. $Q_e=0.1$ is a normalized factor in
association with hydrodynamical simulation results. Moreover, the angular momentum
exchange between the embryos and the gas disk will result in orbital migration of planets.
When the planets are less massive, they will undergo type I migration,
whereas they grow large enough, they will experience type II migration \citep{idalin04}.

The timescale of type I migration can be assessed using linear
model and the net loss on embryo \citep{GT79,Ward97,Tan02} is
expressed as

\begin{eqnarray}
\tau_{\rm migI}=\frac{a}{|\dot{a}|}=\frac{1}{(2.7+1.1\beta )}
\left(\frac{M_*}{m}\right)\left(\frac{M_*}{\Sigma_ga^2}\right)
\nonumber\\
\times\left(\frac{h}{a}\right)^2
\left[\frac{1+(\frac{er}{1.3h})^5}{1-(\frac{er}{1.1h})^4}\right]\Omega^{-1},~~~~~~
\label{tauI}
\end{eqnarray}
where $e,~r,~h$, and $\Omega$ are the same meaning as given in
Equation \ref{damp}. Using the gas density profile in Equation
(\ref{dens}), we can achieve $\beta =-d\ln \Sigma_g/d\ln a=1.5$.

Recent work showed that type I migration could be directed
either inward or outward depending on different planetary
and gas disk properties \citep{cos13}, but in general,
outward type I migration can simply at work when the mass of the planets
can be several $M_\oplus$ \citep{cos14, lega14}, which are much larger than the embryos and planetesimals in our model,
so we do not consider outward type I migration in this work.

The planet will experience type II migration when a planet grows to a massive one ($M\geq M_{\rm crit}$)  in the
viscous disk \citep{lp93}. The timescale of type II migration is
described as \citep{IL08}

\begin{eqnarray}
\tau_{\rm migII1}\simeq 5\times
10^5f_g^{-1}~~~~~~~~~~~~~~~~~~~~~~~~~~~~~~~~~~~~~~~~~~~~~~~~~~~~~~~
\nonumber\\
\times\left(\frac{C_2\alpha}{10^{-4}}\right)^{-1}\left(\frac{m}{M_J}\right)\left(\frac{a}{\rm{1AU}}\right)^{1/2}\left(\frac{M_*}{M_\odot}\right)^{-1/2}
,~~~~~~
\nonumber\\
\tau_{\rm migII2}=\frac{a}{|\dot{a}|}=0.7\times
10^5\left(\frac{\alpha}{10^{-3}}\right)^{-1}\left(\frac{a}{\rm
{1AU}}\right)\left(\frac{M_*}{M_\odot}\right)^{-1/2}, \label{tpii}
\end{eqnarray}
where $\alpha$ and $C_2$ are efficiency factor of angular momentum
transport and reduce factor. Herein $C_2\alpha=10^{-4}$. $\tau_{\rm
migII1}$ is fit to the case that the planet mass is comparable to
the entire mass of the gas disk. While the planet mass is lower than
the mass of the gas disk, $\tau_{\rm migII2}$ is the right fit.
We adopt an empirical formula for the eccentricity damping
from the gas disk, $({\tau_{\rm damp2}})^{-1}=(\dot e/e)=-K |\dot
a/a|$ \citep{lee02}. Herein, we choose K=10.  The critical
mass of planet that can produce a gap is given as

\begin {equation}
M_{\rm{crit}}\simeq
30\left(\frac{\alpha}{10^{-3}}\right)\left(\frac{a}{\rm
1AU}\right)^{1/2}\left(\frac{M_*}{M_\odot}\right)M_\oplus. \label
{cri}
\end {equation}

In this work, besides mutual gravitational interaction among the objects in the system,
we also consider the orbital migration, eccentricity damping and aerodynamical drag
of planetary embryo (planetesimal). The acceleration of
the planet or planetary embryo (planetesimal) with mass $m_i$ is expressed as

\begin{eqnarray}
\frac{d}{dt}\textbf{V}_i =
 -\frac{G(M_*+m_i
)}{{r_i}^2}\left(\frac{\textbf{r}_i}{r_i}\right) +\sum _{j\neq i}^N
Gm_j \left[\frac{(\textbf{r}_j-\textbf{r}_i
)}{|\textbf{r}_j-\textbf{r}_i|^3}- \frac{\textbf{r}_j}{r_j^3}\right]
\nonumber\\
+\left\{
\begin{array}{ll}
\textbf{F}_{\rm damp1}+\textbf{F}_{\rm aero}+\textbf{F}_{\rm migI} \rm~(for~planetesimals/embryos)\\
\textbf{F}_{\rm damp2}+\textbf{F}_{\rm migII}
\rm~~~~~~~~~~~~~(for~giant~planets)
\end{array}
\right. \label{eqf}
\end{eqnarray}
where

\begin {eqnarray}
\begin{array}{lll}
\textbf{F}_{\rm damp_{1,2}} = -2\frac{\displaystyle (\textbf{V}_i \cdot
\textbf{r}_i)\textbf{r}_i}{\displaystyle r_i^2\tau_{damp_{1,2}}},
\\
\cr\noalign{\vskip 0.5 mm} \textbf{F}_{\rm
migI}=-\frac{\displaystyle \textbf{V}_i}{\displaystyle 2\tau_{\rm
migI}},
\\
\cr\noalign{\vskip 0.5 mm} \textbf{F}_{\rm migII} =
-\frac{\displaystyle {\bf V}_i}{\displaystyle 2 \tau_{\rm migII}},
\end{array}
\end{eqnarray}
where $\textbf{r}_i$ and $\textbf{V}_i$ represent the position and
velocity vectors of planet $m_i$ and all vectors are given in
stellar-centric coordinates.

\section{Numerical Simulations and Results}

\subsection {Initial conditions}

In our simulations, all bodies are assumed to be initially in
coplanar and near-circular trajectory orbiting the central star,
where the argument of pericenter, mean anomaly, and longitude of
ascending node are randomly distributed between\textbf{ $0^{\circ}$ to
$360^{\circ}$}, respectively. In total, five runs were carried out
for the investigation of planetary formation.

In simulation S1-S5, the planetary system is composed of the host
star, one or two giant planets, and 2000 equal-mass planetesimals.
As shown in Table \ref{tb1} and Equation (8), each planetesimal
performs aerodynamical drag, eccentricity damping, Type I migration
and the gravitational perturbations from Jupiter or Saturn and other
planetesimals in the system. The giant planets are modeled to suffer
from eccentricity damping, Type II migration, and the interaction
from the planets and small bodies. In several runs, the giant
planets are assumed to be fixed about the original region without
migration. In the following, we will briefly summarize each case.

\begin{table*}
\centering \caption{The initial conditions of the runs.
The meaning of Force is given by Equation 8.
%\vspace{0.5cm}
  \label{tb1}}
\begin{tabular*}{14cm}{@{\extracolsep{\fill}}llcccc}
\hline
Name& Mass of planetesimal & No.  & Jupiter &Saturn &Force\\
&$M_\oplus$& Planetesimals & Yes or Not & Yes or Not & on giant planets\\
\hline
S1&$5\times 10^{-3}$  &2000 &Y &N & $F_{\rm migII}$\\
S2&$5\times 10^{-3}$  &2000 &Y &N & No migration\\
S3&$5\times 10^{-3}$  &2000 &Y &Y & No migration\\
S4&$5\times 10^{-3}$  &2000 &Y &Y & $F_{\rm migII}$ \\
S5&$5\times 10^{-3}$  &2000 &Y &Y & $F_{\rm migII}$ \\
\hline
\end{tabular*}
\end{table*}

\begin{table*}
\centering \caption{The statistics of the final destination of the planetesimals in each runs.
%\vspace{0.5cm}
  \label{tb2}}
\begin{tabular*}{14cm}{@{\extracolsep{\fill}}llcccc}
\hline
Name& Time & Percentage of mass  & Percentage of Mass &Saturn &Force\\
& Myr & (inside 0.1 AU) & (beyond 5 AU) & Yes or Not & on giant planets\\
\hline
S1&$ 5 $  & 79.5\% &8\% &N & $F_{\rm migII}$\\
S2&$ 3.9 $  &14.1\% &0 &N & No migration\\
S3&$ 3.7 $  &17.9\% &0 &Y & No migration\\
S4&$ 5 $  &50.5\% &14.8\% &Y & $F_{\rm migII}$ \\
S5&$ 5 $  &36.9\% &5.9\% &Y & $F_{\rm migII}$ \\
\hline
\end{tabular*}
\end{table*}

In these simulations, we mainly explore the terrestrial planetary formation
under the circumstance of the existence of giant planets. Thus, in
each simulation we start with a swarm of planetesimals, together
with one or two giant planets in the system, orbiting the central
star. The initial positions for Jupiter and Saturn are set to be 5.0
AU and 9.54 AU, respectively. Originally, the inner region of the
system consists of 2000 planetesimals that each owns a mass of
$5\times 10^{-3}~M_\oplus$, thereby leading to an entire mass of the
planetesimal disk of 10 $M_\oplus$. Herein for all cases, the
planetesimals are initially distributed in the region [0.5, 3.78]
AU.

S1: In this scenario, the planetary system comprises the host star,
the planetesimals, and  one Jupiter-mass planet. The model assumes
that the giant planet suffers from type II migration over the
dynamical evolution.

S2: The initials parameters for Jupiter and planetesimals are the
same as those given in S1. As a comparison,  in this simulation the
model does not account for type II migration for Jupiter.

S3, S4 and S5: in this three runs, the initial system consists of
the host star, the planetesimals, two giant planets -- Jupiter and
Saturn. In simulation S3, both Jupiter and Saturn do not migrate
inward but remain stable orbits. In simulation S4 and S5, both
Jupiter and Saturn are allowed to migrate inward in the dynamical
evolution.

We integrate Equation (\ref{eqf}) using a hybrid symmetric
algorithm in MERCURY package \citep{Cham99}. However, we have
modified codes to incorporate gas drag and orbital migration
scenarios for our simulation. In these runs, mutual interactions of
all bodies are fully taken into account. Two bodies are considered
to be in collision stage whenever their distance is less than the
sum of two physical radii \citep{Cham99}. If two objects collide
each other, they can merge and form a single larger body without any
fragmentation. In our simulation, each run is integrated for 2 - 100
Myr with a time step of 2 days and a Bulirsch-Stoer tolerance of
$10^{-12}$. As usual, when the simulation ends up, the variations of
energy and angular momenta are $10^{-3}$ and $10^{-11}$,
respectively.

\begin{table*}
\centering
\begin{minipage}{160mm}
\caption{Physical parameters of the 3 runs that form 4:2:1 MMR at 100 Myr and comparison with exoplanet systems.
%\vspace{0.5cm}
 \label{tb3}}
\begin{tabular*}{14cm}{@{\extracolsep{\fill}}llllllllllll}

\hline
Name       & Mass & Radius & Period & $a$  & $e$   &\vline~Name & Mass \footnote{Kepler planetary masses estimated with Eq. (1) in \citep{lis11b}.} & Radius & Period & $a$  & $e$ \\
    &$M_\oplus$&$R_\oplus$ & day    & AU   &       &\vline~  &$M_\oplus$ &$R_\oplus$ & day    & AU   &     \\
\hline
S1 Planet 1  & 5.48 & 2.47 & 6.62  & 0.069 & 0.001 &\vline~Kepler-238 c & 6.0 & 2.39 & 6.16  & 0.069 & -   \\
S1 Planet 2  & 1.62 & 1.65 & 12.25 & 0.104 & 0.038 &\vline~Kepler-238 d & 10  & 3.07 & 13.23 & 0.115 & -   \\
S1 Jupiter & 333.55 & 9.72 & 24.91 & 0.167 & 0.000 &\vline~Kepler-238 e & 77  & 8.26 & 23.65 & 0.169 & -   \\
\hline
S4 Saturn  & 95.69  & 6.41 & 5.64  & 0.062 & 0.001 &\vline~Kepler-302 b & 18  & 4.06 & 30.18 & 0.193 & -   \\
S4 Planet 1  & 5.05 & 2.41 & 11.72 & 0.101 & 0.012 &\vline~-            & -   & -    & -     & -     & -   \\
S4 Jupiter & 333.46 & 9.72 & 24.91 & 0.167 & 0.001 &\vline~Kepler-302 c & 180 & 12.45& 127.28& 0.503 & -   \\
\hline
S5 Planet 1  & 3.69 & 2.16 &  10.02 & 0.0911 & 0.142 &   \\
S5 Jupiter & 333.02 & 9.71 &  20.15 & 0.1449 & 0.025 &   \\
S5 Saturn  & 95.12 &  6.40 &  40.79 & 0.2315 & 0.014 &    \\
\hline
\end{tabular*}
\end{minipage}
%\tablecomments{0.86\textwidth}{*Kepler planetary masses estimated with Eq. (1) in \citep{lis11b}.}
\end{table*}

\subsection {Terrestrial planets formation with giant planets}

The simulations of S1-S5 exhibit classical terrestrial planetary
accretion scenarios in their late stage formation
\citep{cham01,raym04,raym06,Fogg05,Fogg09}.

\subsubsection {Terrestrial planets formation with one giant planet}

\begin{figure*}
   \centering
   \includegraphics[width=\textwidth, angle=270]{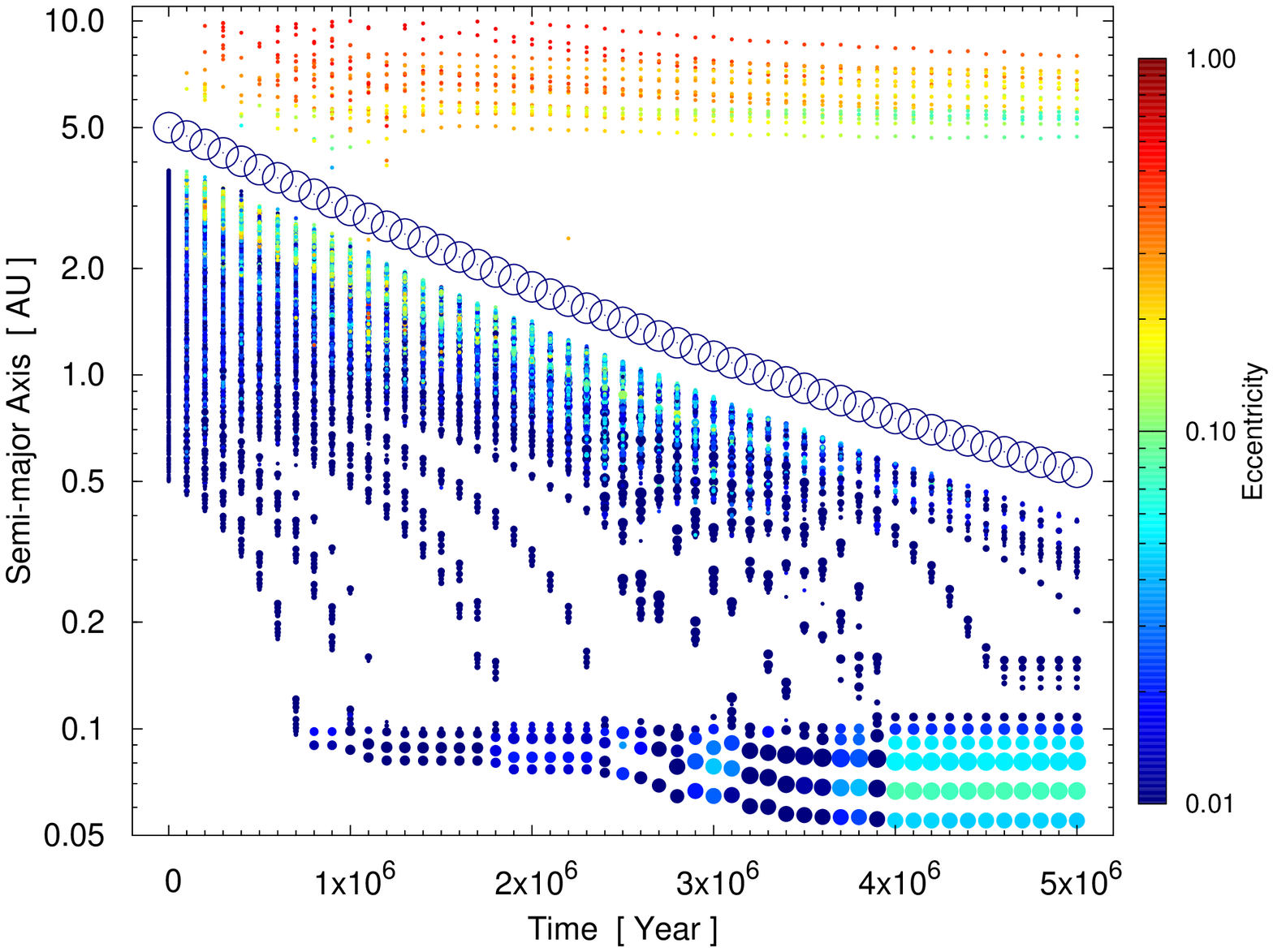}
%   \begin{minipage}[]{170mm}
   \caption{Orbital evolution of the planetesimals and Jupiter mass planet in simulation S1.
The solid points show the planetesimals.
The radii of the planetesimals are proportional to their mass. The
color of each point shows the orbital eccentricity. The open circle
shows the giant planet. As the Jupiter mass planet migrating inward,
the planetesimals at the MMR orbits were excited to high eccentric
orbits. Due to the inward-migrating giant planet, planetesimals were
shepherded inward and hence the accretion rate in the inner part of
the planetesimal disk increases. There are also some planetesimals
were scattered to the outer part of the disk. Several terrestrial
planets form at 5\,Myr. }
%  \end{minipage}
   \label{Fig1}
   \end{figure*}

Figure~\ref{Fig1} shows the orbital evolution of the planetesimals
and the Jupiter-mass planet in the first 5 Myr for simulation S1. At
an early time, the Jupiter-mass planet migrates inward in the disk,
thereby giving rise to the planetesimals nearby the giant planet to
either be scattered outward into high-eccentric orbits \citep{ms03}
or shepherded inward by the giant planet's moving mean motion
resonances \citep{Tan99,Fogg05}. The buildup of inner material
induces rapid growth of two close-in planets within a few Myr. At 5
Myr, five terrestrial planets form inside 0.1 AU. The total mass of
these five planets is 7.95 $M_{\oplus}$, corresponding to $\sim
80\%$ of the initial building materials set in the simulation. There
are 16 planetesimals that were scattered to orbits beyond 5 AU, the
total mass of them is $\sim$ 0.08 $M_{\oplus}$, the final
destination of the planetesimals is shown in Table \ref{tb2}. The
simulation S1 continues evolving for 100 Myr. By the end of 100 Myr,
most of the planetesimal in the initial disk have been nearly
cleared up by ejection or collision scenarios, arising from frequent
orbital crossings over the chaotic evolution. We observe that there
also exist a couple of moderate-eccentric planetesimals which are
not involved in accretion process beyond $\sim$ 5 AU for the
remaining disk.

\begin{figure*}
   \centering
   \includegraphics[width=\textwidth, angle=0]{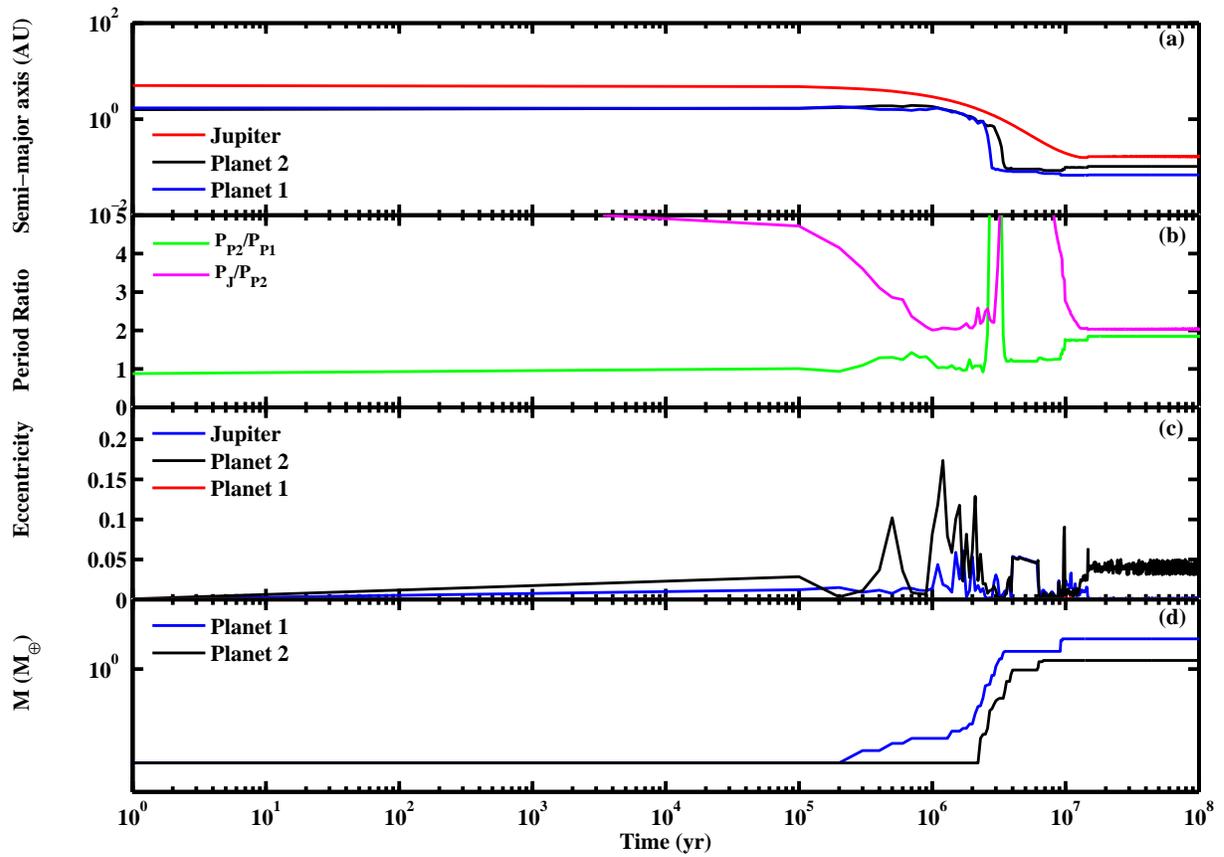}
%   \begin{minipage}[]{170mm}
      \caption{Results of the evolution of semi-major axis (upper panel) and eccentricities
      (lower panel) for simulation S1. Two inner planets undergo type I migration
      while the outer one under the influence of type II migration.
      Note that three planets are close to a 4:2:1 MMR.}
%   \end{minipage}
   \label{Fig2}
   \end{figure*}

Moreover, we find that two terrestrial planets with a mass of $5.48~M_\oplus$  and
$1.62~M_\oplus$ are eventual survivors in the system, locating at
0.069 AU and 0.104 AU, respectively. These two terrestrial planets
and the Jupiter form a 4:2:1 MMR orbital configuration at $\sim$ 10 Myr,
as shown in Figure~\ref{Fig2}. The discovery of the exoplanets shows that
the planetary systems in the universe are quite diverse. Our simulations present
very interesting results, which may provide some clues to future observations
for such systems.

Our simulation S1 is similar with the work in \citet{raym06} that
investigated habitable terrestrial planet formation under a
migrating giant planet. In their model, they adopted a higher mass
of 17 $M_{\oplus}$ and a more extended planetesimal disk that
arranges from 0.25 to 10 AU. They found that hot Earths can form
interior to a migrating giant due to the shepherding effect, and in
some cases water-rich earth-mass planet can form outside the
migrating giant, located inside the habitable zone. Simulation S1 also
shows similar shepherd mechanism, as shown in Figure~\ref{Fig1}.
Interestingly, the two inner terrestrial planets, as well as the migrating giant
planet, are discovered to finally form a 4:2:1 MMR orbital configuration.
However, we do not observe material around the habitable zone.
In simulation S1, the massive giant planet sweeps away or accrete
most of the materials along its pathway when it migrates inward.
Furthermore, we point out that such material depletion of migrating giant
planet can be observed in our simulation S5 (Figure~\ref{Fig7}),
in which the system contains two migrating giants.

\begin{figure*}
   \centering
   \includegraphics[width=\textwidth, angle=270]{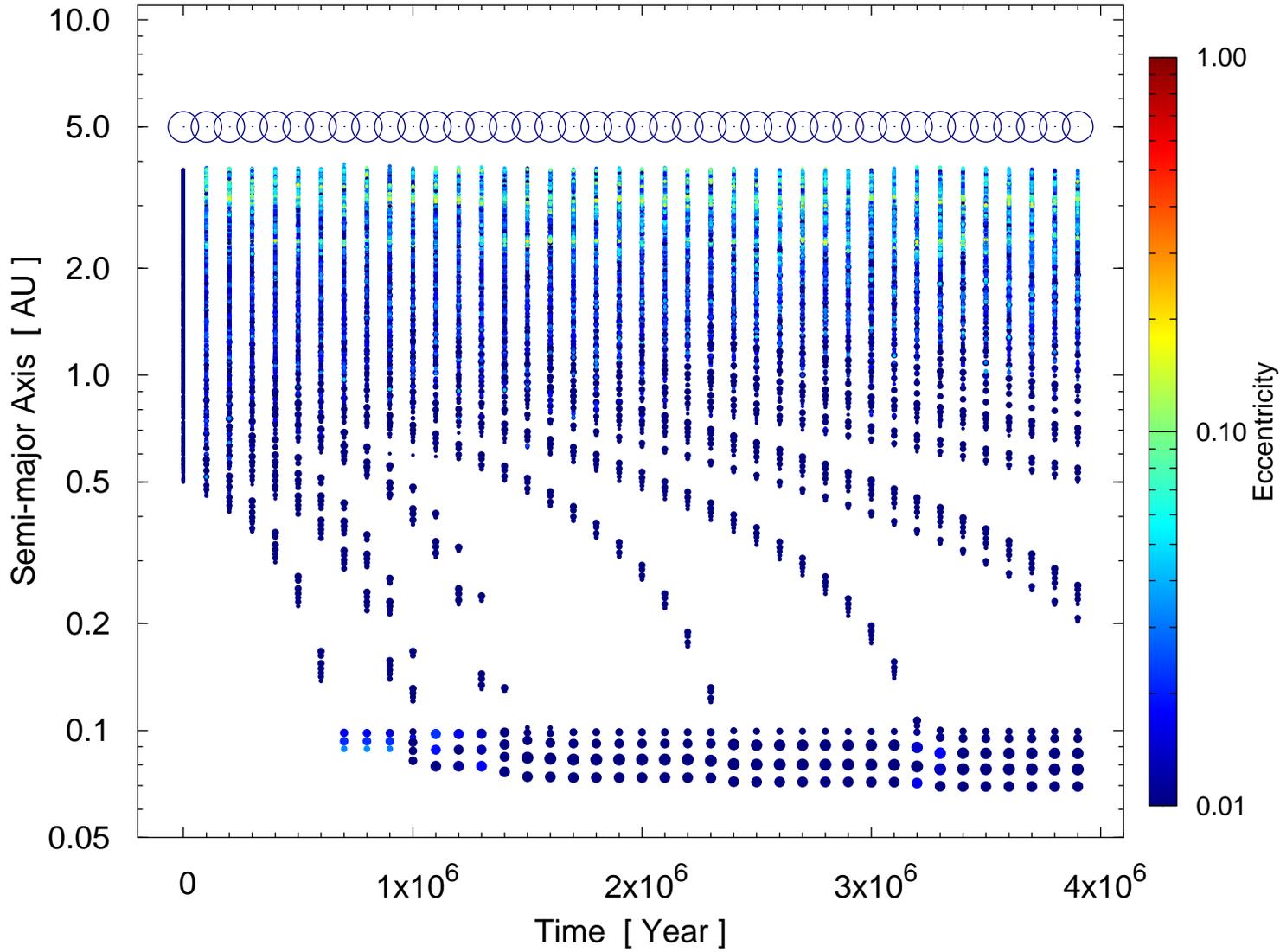}
%   \begin{minipage}[]{170mm}
   \caption{Orbital evolution of the planetesimals and Jupiter mass planet in simulation S2.
  The Jupiter mass planet in this simulation was fixed at 5 AU.
The radii of the planetesimals are proportional to their mass. The
color of each point shows the orbital eccentricity. We can see the
planetesimals at its MMR orbits were excited to high eccentric
orbits. The accretion rate in this simulation is much lower than in
simulation S1. }
%  \end{minipage}
   \label{Fig3}
   \end{figure*}

Initials in simulation S2 are the same as those in S1, however in
this model we simply do not let Jupiter undergo type II migration.
Figure~\ref{Fig3} shows the orbital evolution of the planetesimals
and the Jupiter-mass planet for simulation S2. In similar case, the
planetesimals are quickly excited due to their mutual gravitation,
along with that of the Jupiter-mass planet locating at 5.0 AU. The
objects, involved in the 3:1 MMR at 2.50 AU, 2:1 MMR  at 3.28 AU,
and the 5:3 MMR at 3.70 AU, are stirred within 0.1 Myr. This trend
can be clearly seen by the rise of the eccentricities of the
planetesimals at these locations as shown in Figure~\ref{Fig3}. All
stuff in the planetesimal disk exterior to the 3:2 resonance at 3.97
AU is quickly removed from the system via collision and ejection
resulting from the giant planet. In the time evolution, the
planetesimals' eccentricities can increase and the system becomes
chaotic when the eccentricities are larger enough. The bodies in the
inner disk begin to grow via accretionary collisions within 1 Myr.
The larger bodies tend to have smaller eccentricities and
inclinations, due to the dissipative effects of dynamical friction.
At 3.9 Myr, there are four less massive terrestrial planets (as
compared to S1) with a mass of $0.15 - 0.51 M_\oplus$ formed in the
region [0.07, 0.095] AU. The entire mass of four terrestrial planets
is  $1.41~M_\oplus$, corresponding to $\sim 14.1\%$ of the total
initial materials as shown in Table \ref{tb2}.

In comparison, we can see that an inward-migrating giant planet can
significantly increase the accretion rate in the inner part of
planetesimal disk by the shepherd effect. Meanwhile, it can scatter
the bodies in the inner disk to the outer disk.

\subsubsection {Terrestrial planets formation with two giant planets}

In the following, we will present the simulation outcomes of the
terrestrial planetary formation co-existence with two giant planets
in the planetary systems.

\begin{figure*}
   \centering
   \includegraphics[width=\textwidth, angle=270]{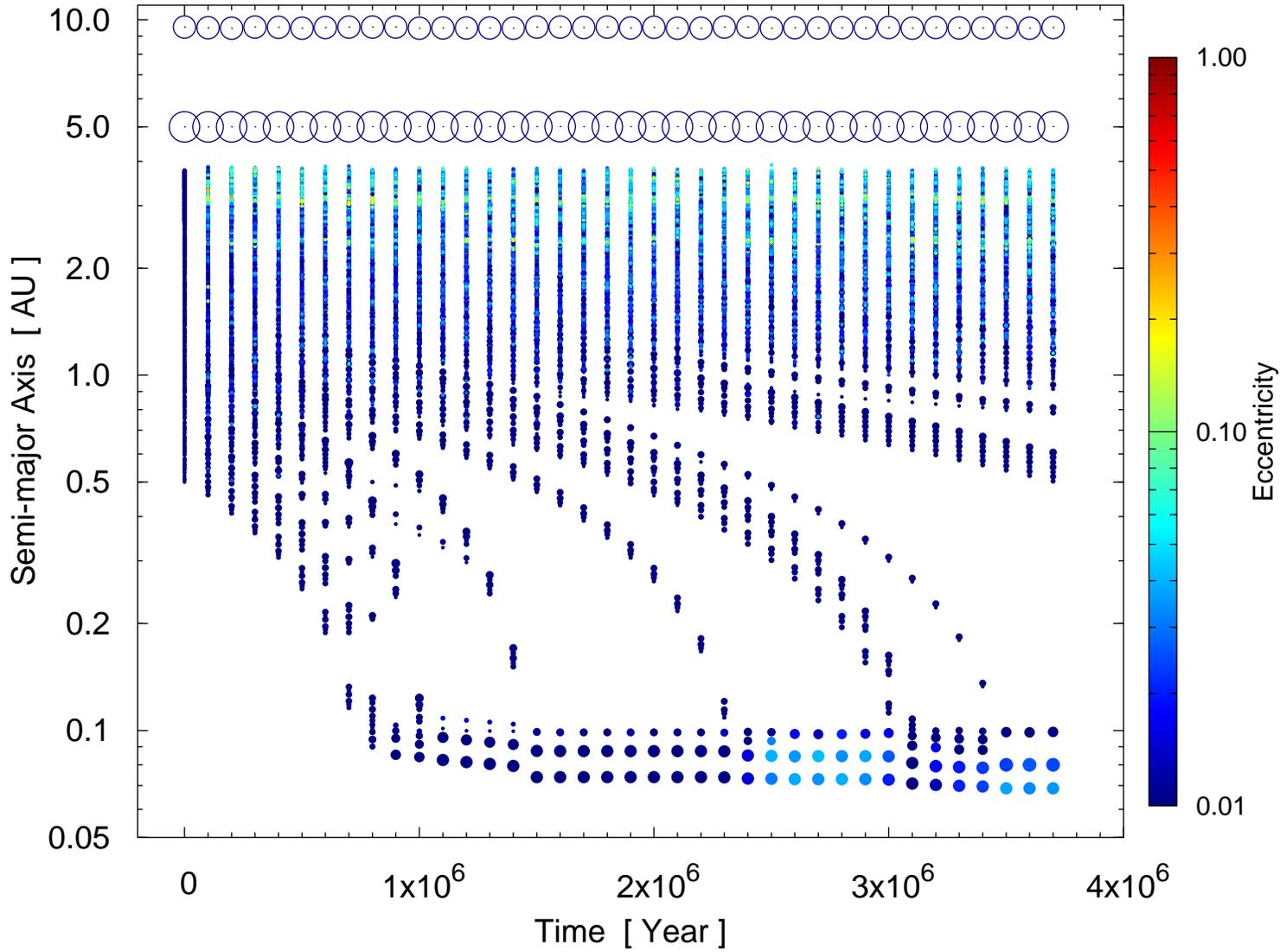}
%   \begin{minipage}[]{170mm}
   \caption{Orbital evolution of the planetesimals and two giant planets in simulation S3.
  The Jupiter and Saturn in this simulation was fixed at 5 and 9.2 AU.
The radii of the planetesimals are proportional to their mass. The
color of each point shows the orbital eccentricity. The
planetesimals at the MMR orbits of the Jupiter mass planet were
excited to high eccentric orbits. }
%   \end{minipage}
   \label{Fig4}
   \end{figure*}

In simulation S3, two giant planets, which bear a Jupiter or Saturn
mass, respectively, do not perform type II migration in the
simulation. Therefore, similar to simulation S2, the planetesimals
in the disk can be swiftly excited due to gravitational
perturbations from two giant planets. In particular, the bodies,
which are involved in the MMR with the Jupiter-mass planet, then
acquire moderate eccentricities within 0.1 Myr. As compared to
Jupiter, the Saturn-mass planet plays a less dominant role in the
mass accretion of planetesimals in forming terrestrial planets. In
the meanwhile, owing to the co-existence of two giant planets, the
terrestrial formation process can be speeded up although Jupiter and
Saturn do not deviate much from their initial orbits in this run.
Figure~\ref{Fig4} shows temporal orbital evolution of all the
planetesimals and the giant planets in simulation S3. At $\sim$ 2
Myr, three terrestrial planets are yielded with a mass in the range
$0.10 - 0.57 M_\oplus$, and they orbit in the broad region from 0.07
AU to 1 AU.  At the time of 3.7 Myr, the entire mass of four terrestrial
planets formed inside 0.1 AU is $1.79~M_\oplus$, occupying $17.9\%$ of
the initial mass of the planetesimal disk. Similar with the simulation S2,
there is no planetesimal beyond 5 AU.

In order to investigate the role of type II migration, we have
performed two additional runs for the systems composed of two giant
planets, Simulation S4 and S5. In this two runs, both Jupiter and
Saturn undergo type II migration. Our results provide evidence that
terrestrial formation may take place in the inner region of the
planetary systems (\citet{Fogg05,man07,Fogg09}). However, new runs
further show some interesting outcomes - the 4:2:1 MMR configurations
are formed for these systems.

\begin{figure*}
   \centering
   \includegraphics[width=\textwidth, angle=270]{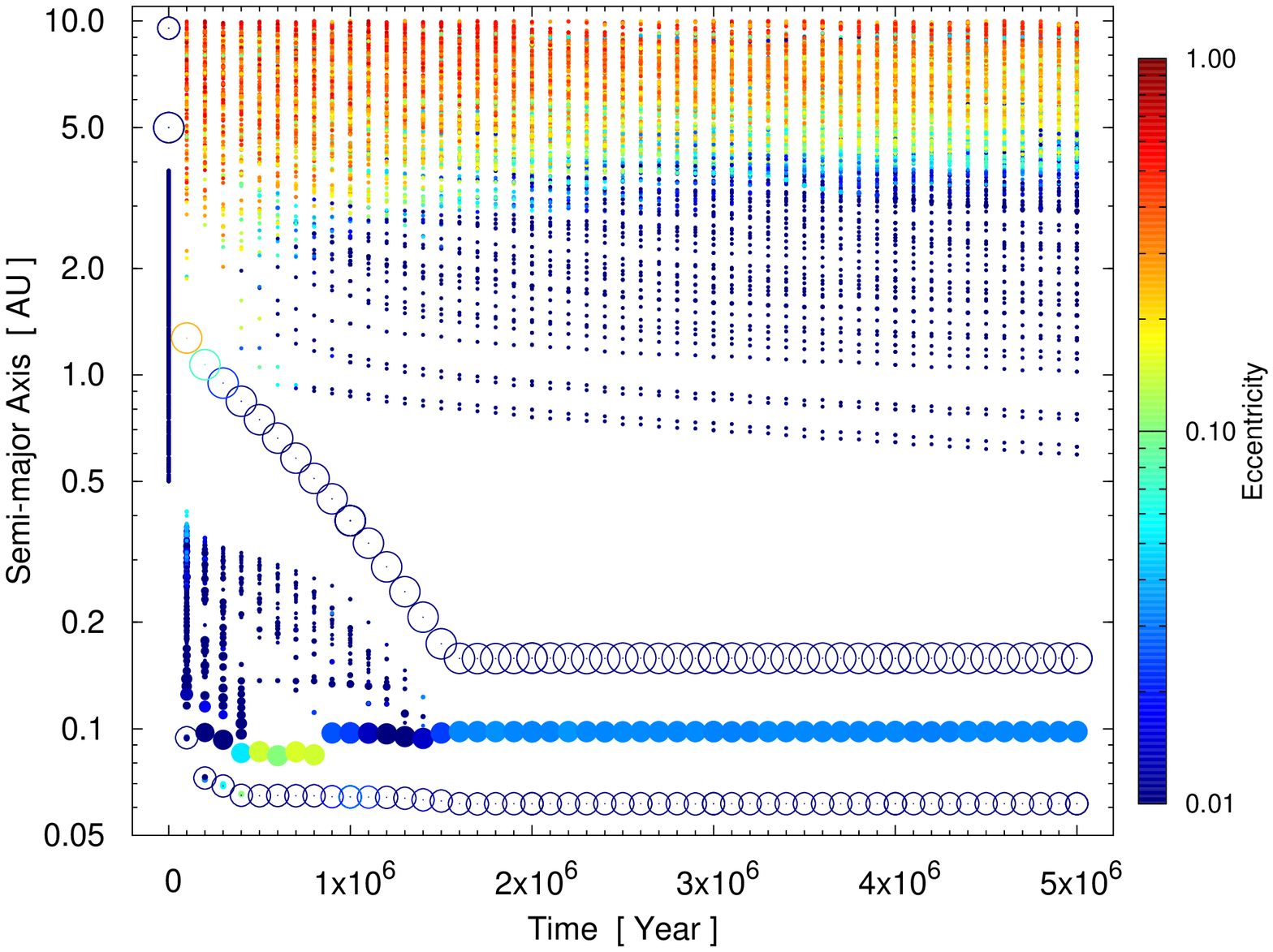}
%   \begin{minipage}[]{170mm}
   \caption{Orbital evolution of the planetesimals and two giant planets in simulation S4.
  The Jupiter and Saturn in this simulation underwent type II migration.
The radii of the planetesimals are proportional to their mass. The
color of each point shows the orbital eccentricity.
      Note that three planets are close to a 4:2:1 MMR.}
%   \end{minipage}
   \label{Fig5}
   \end{figure*}

Figure~\ref{Fig5} shows temporal orbital evolution of all the
planetesimals and giant planets. According to Equation
\ref{tpii}, $\tau_{\rm migII1}$ is proportional to the planetary
mass, thus we learn that Saturn migrates faster than Jupiter does,
indicating that there would be a possibility for two giant planets
to rendezvous in the system. At 0.02 Myr, Jupiter's eccentricity is
gradually pumped up to 0.30 due to gravitational interaction from
Saturn when it moves inward closer to the Jovian planet.

\begin{figure*}
   \centering
   \includegraphics[width=\textwidth, angle=0]{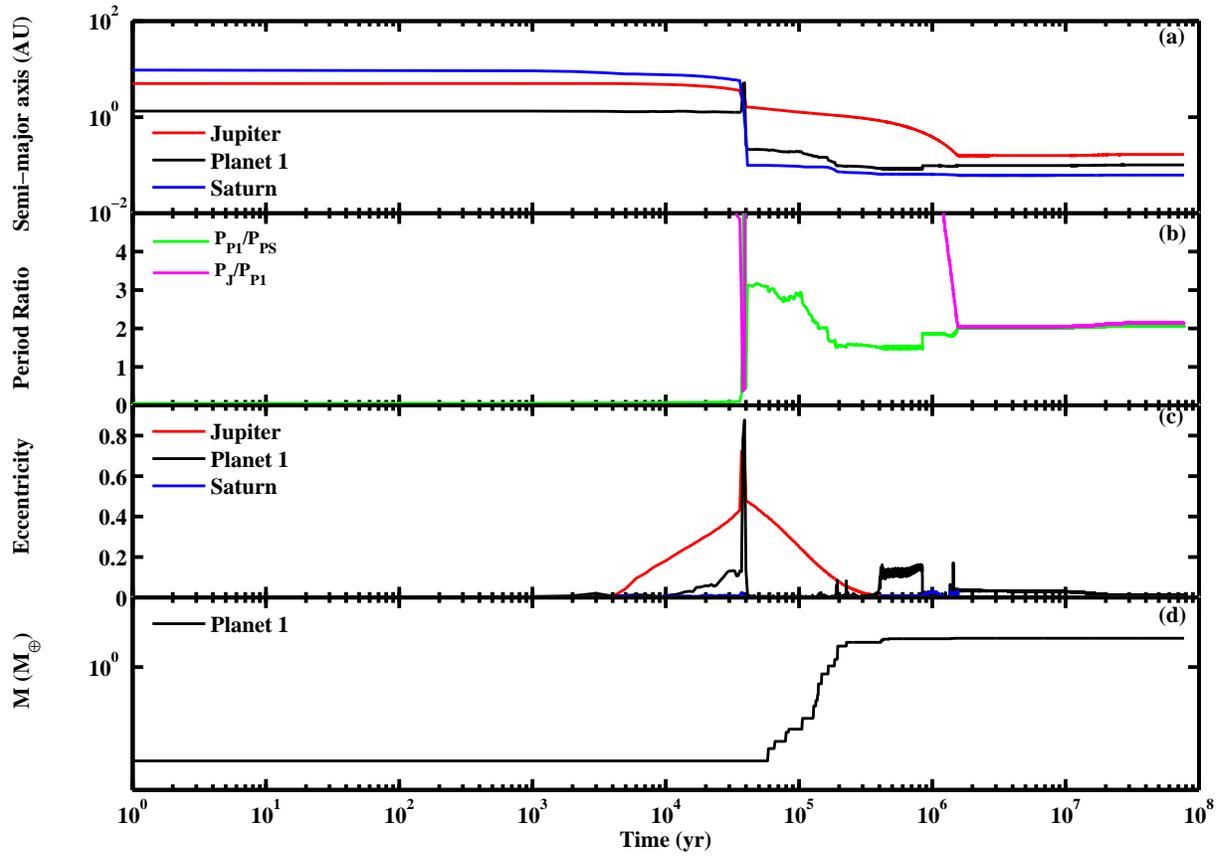}
%   \begin{minipage}[]{170mm}
      \caption{Results of the evolution of semi-major axis (upper panel) and eccentricities
      (lower panel) for simulation S4. One inner planet undergo type I migration
      while the outer two under the influence of type II migration.
      Note that Saturn is kicked inward and three planets are evolved into 4:2:1 MMR.}
%   \end{minipage}
   \label{Fig6}
   \end{figure*}

\begin{figure*}
   \centering
   \includegraphics[width=\textwidth, angle=270]{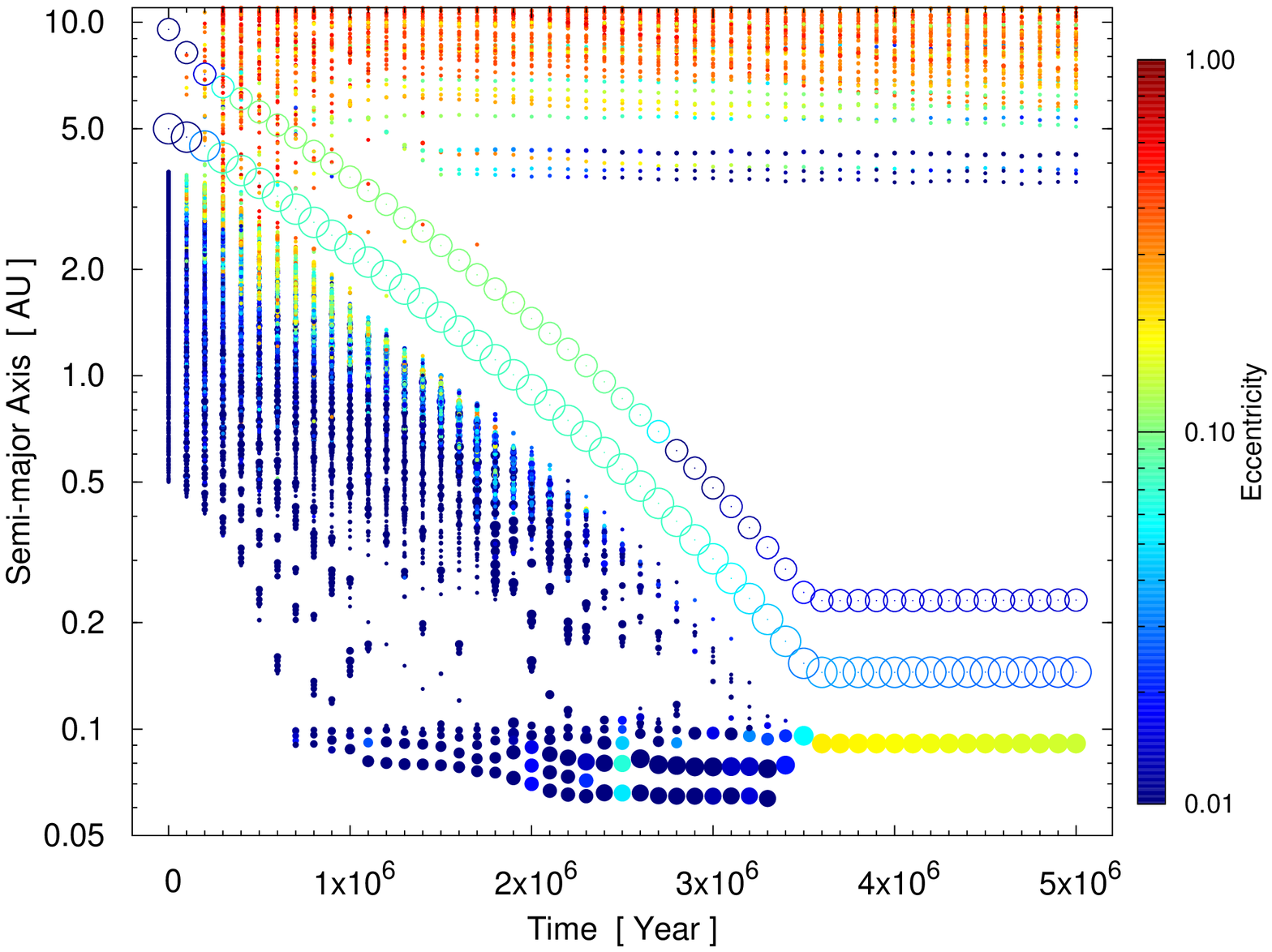}
%   \begin{minipage}[]{170mm}
   \caption{Orbital evolution of the planetesimals and two giant planets in simulation S5.
  The Jupiter and Saturn in this simulation underwent type II migration.
The radii of the planetesimals are proportional to their mass. The
color of each point shows the orbital eccentricity.
      Note that three planets are evolved into 4:2:1 MMR.}
%   \end{minipage}
   \label{Fig7}
   \end{figure*}

At 0.036 Myr, Jupiter and Saturn reach the orbit at 5.75 AU and 3.58
AU, respectively, indicating that they pass through a 1:2 mean
motion resonance. This resonance crossings may have excited the
orbital eccentricities of the planets which cross the resonance
\citep[e.g.,][]{tsiganis05,morb07}. Thus, the orbital crossing between
two giant planets happens within several thousand years and brings
about fairly chaotic behaviors for them. Hence, a sudden jump in the
Jupiter's eccentricity occurs up to $\sim$ 0.72 at 0.037 Myr,
whereas Saturn may obtain a moderate eccentricity right after
the close encounter but its eccentricity is quickly damped by the gas disk.
From the simulations, we find that Jupiter experienced close encounters
with Saturn at the maximum star-centric distance at $\sim$ 2.41 AU,
which leads to strong interaction between them. From $\sim$ 0.037
Myr to 0.04 Myr, note that the semi-major axis of Jupiter drops down
from 3.58 AU  to 1.63 AU, whereas that of Saturn dramatically
decreases from 5.75 AU  to 0.45 AU.  Furthermore, such chaotic
motions for Jupiter and Saturn trigger catastrophic fate for the
planetesimals in the system, where most of their orbits are severely
excited and scattered.  As Figure~\ref{Fig6} shown, the eccentricity
of the planetesimal (Planet 1) is first kicked to be 0.88 then falls
to 0.13, while its semi-major axis can suddenly increase amount up
to 5.18 AU, then soon drop down to 0.26 AU over the time span.

\begin{figure*}
   \centering
   \includegraphics[width=\textwidth, angle=0]{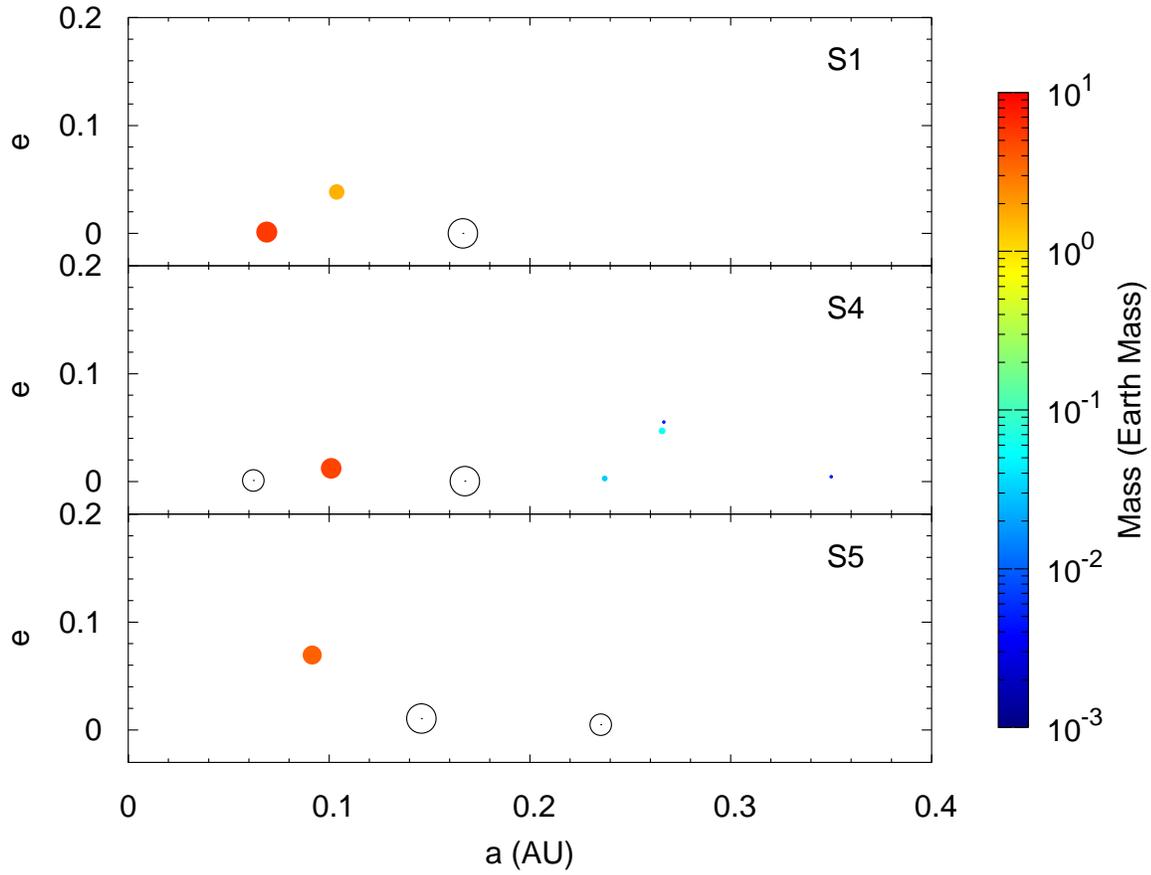}
%   \begin{minipage}[]{170mm}
   \caption{Final configuration of the 3 runs, S1, S4 and S5, that have formed 4:2:1 MMR at 100 Myr. The open circle show the giant planets. The solid point show the formed terrestrial planets. The color of the terrestrial planets show their masses.}
%  \end{minipage}
   \label{Fig8}
   \end{figure*}

As the two giant planets continue migrating inward, the 2:1 MMR
between them is broken up. Subsequently, Saturn is kicked into the
region $\sim$ 0.1 AU due to dynamical instabilities. By 10 Myr,
the giant planets have moved very close to their central star. In
the late stage, the eccentricity of Jupiter is gradually damped by
dynamical dissipation. However, an inner terrestrial planet (Planet
1) is excited onto a highly eccentric orbit exterior to the regime
of two giant planets, and soon crosses Jupiter's orbit, then finally
captured inside the orbits of Jupiter and Saturn. The mass growth
for Planet 1 starts from one planetesimal at $\sim$ 0.06 Myr to a
super Earth with a mass of $5.05~M_\oplus$ at $\sim$ 0.40 Myr.
Finally, Planet 1 remains at 0.101 AU, whereas Saturn and Jupiter
cease migrating at 0.062 AU and 0.167 AU, respectively. This
planetary system shows a very interesting configuration that a super
Earth locates between two giant planets' orbits, and three planets
are trapped into a near 4:2:1 MMR (see Figure~\ref{Fig8}).
Accordingly, the residual materials of disk are either removed
out of the system or scattered into distant orbits.
Their eccentricities rise along with their semi-major axes.

The scenario of S4 shows a resemblance to NICE
Model~\citep[e.g.,][]{tsiganis05,morb07}, which proposes that the
Late Heavy Bombardment (LHB) -- a spike in the impact rate on
multiple solar system bodies that lasted from roughly 400 until 700
Myr after the start of planet formation
~\citep{tera74,cohen00,chapman07} -- was triggered by an instability
in the giant planets' orbits. In the Nice Model, the orbits of the
giant planets would have been in a more compact configuration, with
Jupiter and Saturn interior to the 2:1 resonance. In our model of
simulation S4, the Jupiter-mass and Saturn-mass planets can switch
each place, similar to the case of Uranus and Neptune in the NICE
Model. \citet{ray08b} showed that planet-planet scattering could
create MMRs, most of these resonances are indefinitely stable. The
previous investigations reported that in a convergent migration
scenario for two giant planets in the solar system, the 2:1 MMR
could be formed and then disintegrated in the evolution
\citep{morb07b, zz10}. \citet{zz10b} stated that when Jupiter lies
outside Saturn, convergent migration could occur, Saturn is then
forced to migrate inward by Jupiter where the two planets are
trapped into MMR, and Saturn may move on its tracks of approaching
the central star. Furthermore, to apply our model to explain the
planetary formation of the systems, we extensively examine the
published data by Kepler mission and find a relevant planetary system
(Kepler-302) similar to this simulation. Kepler-302 is a system
consisting of two planets orbiting a star with a mass of 0.97
$M_\odot$, as shown in Table \ref{tb3}. The outer planet of
Kepler-302c is more massive than the inner companion Kepler-302b, and
the two planets have approximate sizes as compared with those of the giant
planets in simulation S4, whereas their orbital distances from the
host star are a bit farther than those of the planets in the
present run. Interestingly, the two giant planets of Kepler-302 are
nearly close to a 4:1 MMR \citep{rowe14}, suggesting that the
system may have gone through similar migration process as in our
simulation. Therefore, our model herein presents a likely
formation scenario for two planets that are close to a 4:1 commensurability,
and also provides evidence that terrestrial planet formation under the
influence of two giant planets in the compact system.

In simulation S5, although Saturn migrates faster than Jupiter,
there is no rendezvous between Saturn and Jupiter. Instead, when
Saturn enters the 2:1 MMR orbit of Jupiter at 0.02 Myr, these two
giant planets were locked in the 2:1 MMR configuration and they
migrate inward together. Figure~\ref{Fig7} shows temporal orbital
evolution of the simulation S5 for the first 5 Myr. Jupiter's
eccentricity is gradually pumped up to 0.30 due to gravitational
interaction from Finally, a terrestrial planet of 3.69 $M_{\oplus}$
form inside 0.1 AU. This new terrestrial planet and the two giant
planets form a 4:2:1 MMR orbital configuration at $\sim$ 5 Myr.
Although Jupiter keeps migrating until ~10 Myr in S1,
Jupiter ceases its migration at 1.5 Myr and 3.5 Myr in S4 and S5,
respectively. This is probably due to the torque from Planet 1, and
the torque depends on the edge.

Unlike in simulation S3, in simulation S4 and S5 we can see
planetesimals beyond 5 AU due to scattering of inward-migrating
giant planets. At 5 Myr, there are about 1.48 $M_{\oplus}$ materials
beyond 5 AU in simulation S4, corresponding to $14.8\%$ of the
initial total mass. While in simulation S5,  there are $\sim$ 0.59
$M_{\oplus}$ beyond 5 AU.

Figure~\ref{Fig8} shows the final configuration in the inner disk
for the three runs that have turned on type II
migration. Interestingly, all these runs yield 4:2:1 MMR. In
simulation S1, two terrestrial planets formed interior to the
Jupiter mass planet. The migrating giant increased the accretion
rate in the inner disk and cleared all material in its pathway.
As a result, there is nothing left between 0.2 to 4.5 AU. In
simulation S4, there was an orbital crossing between the migrating
Jupiter mass and Saturn mass planets, and such a chaotic event
changed the orbital configuration of the two giant planets. Hence,
we note that a close-in terrestrial planet formed between two
giant planets. There are some residual materials between 0.2 to 5 AU
in this run, and this could be a consequence of the quickly orbital
change of the two giants at 0.1 Myr, as shown in Figure~\ref{Fig6}.
While in simulation S5, there is no chaotic orbital change of the
Jupiter mass and Saturn mass planets. Similar to simulation S1,
we see that only one terrestrial planet formed interior to
two migrating giants, and in this case the material of the region
between 0.2 and 4.5 AU of the disk was fully cleared by two giant planets.

\section{The emergence of 4:2:1 MMRs only with terrestrial planets}

The investigations imply that the planetary configurations
in association with a near 4:2:1 MMRs would be common in the
universe. Herein, we consider two planets with a period ratio in the
range of [1.83, 2.18] as a pair near 2:1 MMR \citep{lis11b}. For
example, Jupiter's Galilean moons are known to be locked into a
so-called three-body Laplacian resonance (i.e., 4:2:1 MMR) in our
solar system. On the other hand, other evidences are found in the
exoplanetary systems, in which three planets of the GJ 876 system
\citep{marcy01, rivera10, batygin15, nelson16} and Kepler-79
(KOI-152) \citep{Ste10, wjz12} are reported to be close to 4:2:1
MMR. In this work, we have provided substantial evidences on this
from the simulations. For example, in simulation S1, S4 and S5 of
Section 3, we show that the systems initially composed of one or two
giant planets can finally produce 4:2:1 MMRs, in which the formed
terrestrial planets are involved in. Naturally, a question arises -
what if the systems only contain terrestrial planets?

As mentioned previously, the released Kepler outcomes report that a great number of planetary
systems harboring terrestrial planets involved in a near 2:1 resonance. Furthermore,
a portion of the terrestrial planets discovered by Kepler are in near 4:2:1 resonance.
Herein, we further investigate the configuration formation of those systems only with terrestrial planets
to understand whether the 4:2:1 MMRs can be produced in
such systems. With these purpose, we perform additional runs to explore
the systems that simply consist of terrestrial planets through simulations.

\begin{figure*}
   \centering
   \includegraphics[width=\textwidth, angle=0]{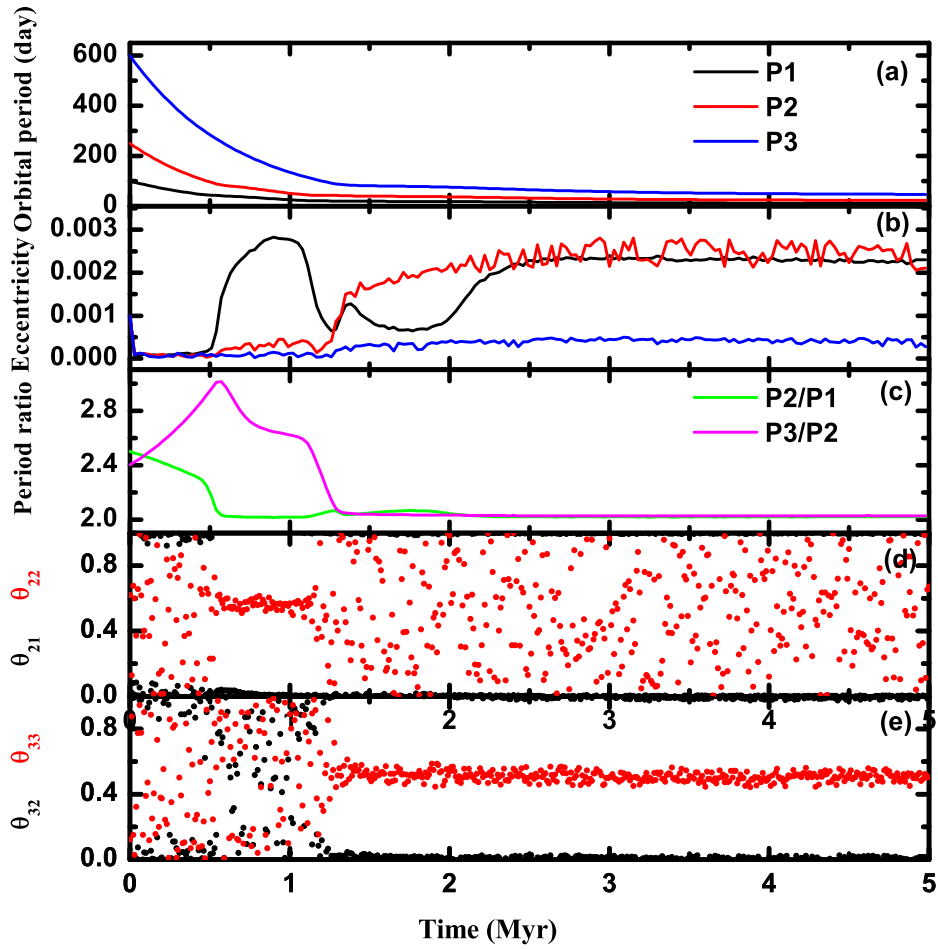}
%   \begin{minipage}[]{170mm}
   \caption{A typical run of simulations for terrestrial planets.
   Panel (a) shows the evolution of the orbital periods.
   Panel (b) displays the evolution of eccentricities and Panel (d) means the period ratio changed with the time.
   In panel (a) and (b), the black, red and blue lines represent the innermost (P1), middle (P2) and the outermost (P3) planets, respectively.
   In panel (c), the green line and pink line represent the period ratio of P2/P1 and P3/P2, respectively. Panel (d) and (e) display the resonance angles of each pairs of
   planets, where
   $\theta_{21}=2\lambda_1-\lambda_2-\varpi_1$, $\theta_{22}=2\lambda_1-\lambda_2-\varpi_2$, $\theta_{32}=2\lambda_2-\lambda_3-\varpi_2$, and $\theta_{33}=2\lambda_2-\lambda_3-\varpi_3$.}
%  \end{minipage}
   \label{Fig9}
   \end{figure*}

In our additional simulations, there are three terrestrial planets
initially with masses less than 30 $M_\oplus$ around central star,
and they are assumed to migrate from the outer region of the system
under the effect of the gas disk. The gas density profile is
proportional to $r^{-1}$, and an empty hole located in the inner
region of the system. During the formation process, the terrestrial
planets suffer from type I migration and gas damping. The timescales
of type I migration and gas damping are shown in Equation
(\ref{damp}) and (\ref{tauI}), respectively. In total, we carry out
seven groups of simulations, and each group contains five runs by
considering different speed of type I migration. We add an
enhance factor $\eta$ to the timescale of type I migration. The five
runs correspond to $\eta$ times of typical timescales of type I
migration as shown in Equation \ref{tauI}. The values of $\eta$ are
100, 33.3, 10, 3.3, and 1, respectively. Figure \ref{Fig9} shows one
of the runs with typical outcomes.Considering various
speed of type I migration, three planets are located at 140, 500,
and 1450 d when $\eta$ is larger than 33.3, otherwise they are
located at 100, 250, and 600 d, respectively. The masses of the planets in
five groups are 5, 10, and 15 $M_\oplus$, respectively, while in
other two groups they are set to be 5, 5, and 5 $M_\oplus$,
respectively. In Figure \ref{Fig9}, the formation of the innermost,
the middle and the outermost planets are marked by the black, red
and blue lines, respectively. Panel (a), (b) and (c) show the
evolution of the orbital periods, eccentricities, and the period
ratios, respectively. Panel (d) and (e) display the 2:1 mean
motion resonance angles. In the case with terrestrial planets in
the system, they all undergo type I migration. Due to the short
timescales for the first two planets, they are trapped into 2:1 MMR
first. From Panel (c) in Figure \ref{Fig9}, we find that the first
and second planets are involved into 2:1 MMR at $\sim$ 0.5 Myr, and
then three terrestrial planets are trapped into 4:2:1 MMRs at $\sim$
1.3 Myr. From Panel (b), we can learn that when the planet pairs
enter into MMRs, their eccentricities are excited, but due to the
damping of gas disk, they cannot be stirred up to high values. The
4:2:1 MMR is usually produced when the first planet is trapped in
the edge of the holes in the gas disk. The formation process is
similar to that mentioned in \citet{pn08}, the MMR formed when the
planet is captured in the gap of the other planet. In addition, we
investigate the resonance angles of each pair which are
shown in Panel (d) and (e) of Figure \ref{Fig9} and find that
$\theta_{21}=2\lambda_1-\lambda_2-\varpi_1$ librates at $\sim
0^\circ$, $\theta_{32}=2\lambda_2-\lambda_3-\varpi_2$ librates at
$\sim 0^\circ$, and $\theta_{33}=2\lambda_2-\lambda_3-\varpi_3$
librates at $\sim 180^\circ$, where the subscript 1 and 2 represent
the orbital elements of Planet 1 and 2, respectively. At the end of
the simulation, three planets locate at about 11.52, 23.26, 47.17
days, respectively. This configuration is similar to the system
Kepler-92 which contains three planets in the orbital periods of
13.75, 26.72, and 49.36 days, respectively. In final, by examining
all runs from seven groups, we find that 6 out of 35 runs (17.1\%)
show three planets are finally involved in 4:2:1 MMRs, whereas 10
runs report that only the two inner planets enter into 2:1 MMR, and
two runs for two inner planets in 3:2 MMR. About half of the
simulations (48.6\%) are not in MMR. Based on our
simulations, the proportion that planets pairs trapped near 4:2:1
MMRs appears to be high. There are several reasons that could lead to such results.
Firstly, we set the initial orbital separation of planets at $\sim$ 20 Hill radius,
which is a little larger than the empirical value of 12 Hill radius
\citep{idalin04}. In such cases, most of the planets are first trapped by the 2:1 MMRs
during their evolution. However, as a comparison, if they are initially in a compact configuration,
the probability of the forming planets in 2:1 MMRs will significantly decrease.
As shown in the work of \cite{wang14}, they found that if the planets pairs were initially  in
compact configuration at the separations of $\sim$ 15 Hill radius,
the probability of 2:1 MMRs decreased to 10\% in 50 runs and there was no simulation
that three planets were evolved into 4:2:1 MMRs. Secondly, in our simulations
we slow down the speed of type I migration for low-mass planets, and this could
further explain why two planets are fairly easier to be trapped into 2:1 MMRs.
Finally, the final orbital periods of the planets in all
simulations are less than 50 days which may be affected by the tidal
force that arises from the central star. The tidal effect may lead to the
deviation from 2:1 MMRs \citep{lee13}. In summary, the above-mentioned
factors may play a major role in causing the planets pairs evolved in near 4:2:1 MMRs
at a relatively moderate possibility. The simulations in \cite{raym06} utilized the
typical speed of type I migration which was very fast, and the
embryos were initially located close to each other in compact configuration.
These may be the leading differences producing diverse results between our simulations and
those of \cite{raym06}.

By comparing S1, S4 and S5 simulation containing giant planets and
the runs that simply consist of terrestrial planets in this section,
we find that the timescales of the planet pairs trapped into MMRs
are various for different cases. For the systems that only contain
terrestrial planets, the two inner planets are found to be trapped
into 2:1 MMRs within 1 Myrs, and the two outer pairs are in
resonance at approximately 2-3 Myrs. Therefore, we may conclude that
three terrestrial planets are in 4:2:1 MMRs within 3 Myrs on
average. In comparison, for the systems consisting of one or two
giant planets, the 4:2:1 MMRs emerges at over 2 Myrs when the
terrestrial planets complete its mass accretion. Additionally, the
eccentricities of planets which is in MMR with giant planets are
always excited to be high values, while the eccentricities of
planets in systems without giant planets will be suppressed to low
values.

\section{Conclusions and Discussions}

In this work, we have investigated terrestrial planetary formation
, especially the configuration formation of a near 4:2:1 MMRs by carrying out
two sets of simulations with and without giant planets. From the
results we find that 4:2:1 MMRs configuration can be formed in the
systems both with and without giant planets. Herein we summarize the
major results as follows.

\begin{enumerate}
  \item In the first sets, we place one or two giant planets into the initial systems as well as
  the planetesimals, to understand terrestrial formation in such cases in S1-S5.

  In simulations S2 and S3, the giant planets do not undergo migrating
  in the disk, and the terrestrial formation procedure can be expedited
  because of the existence of Jupiter-mass and Saturn-mass planets.
  In final, several low-mass close-in terrestrial planets can be formed
  as a result of radial mixing and type I migration. In the two
  runs, there is no 4:2:1 MMRs in the evolution.

  In simulation S1, S4, and S5, we place one or two giant planets in the system
  , which undergo type II migration during the evolution.
  Thus the planetesimals are shepherded inward, cleared or ejected along with
  migration of giant planets.
  The close-in terrestrial planets are yielded with a mass up to several Earth-masses.
  For simulation S1, we find that two short-period terrestrial planets and one giant planet
  are finally involved in 4:2:1 MMR, whereas in the runs of S4, and S5, we show that one
  terrestrial planet and two giant planets enter into the 4:2:1 MMR.
  The resultant results of S1 and S4 are similar to Kepler-238 and 302
  systems in orbital spacing. We examine the mass concentration statistic
  $S_c$ and the orbital spacing statistic $S_s$ \citep{cham01,hansen12} for these systems.
  Comparing the system S1 and Kepler 238, we can note that the orbital spacing statistics are close to each other.
  Their values for two systems are 1.8 and 2.76, respectively, indicating that S1 and Kepler 238 are similar in the orbital distribution.
  However, the mass concentration statistics of them differ from each other, and their values are 125.7 and 30.3, respectively.
  One reason for this difference in mass concentration statistic is that the masses of planets in our simulations are quite uncertain.
  Herein we adopted an estimated masses from Eq. (1) in \cite{lis11b}. If we can obtain the true masses of planets in S1, the difference of $S_c$ could be narrowed down.
  Moreover, we further compare $S_s$ and $S_c$ of the system S4 with those of Kepler 302, where $S_s$ for each system is 5.43 and 4.0, respectively, and $S_c$ for two systems is 10.4 and 23.3, respectively. This may imply that the system S4 and Kepler 302 have similar orbital distribution.
  Therefore, we can conclude that S1 and Kepler 238, S4 and Kepler 302 have similar orbital configurations.

  The present observations show that there is a lack of companion
  planets in the observed hot Jupiter systems. However, from our simulation results,
  we can learn that terrestrial planets can survive inside the inner region of hot Jupiter systems
  based on our formation scenario. As shown in the system S1, the final formed configuration is similar to that in \cite{raym06}.
  Moreover, if the gas disk is small enough, the hot Jupiter could have terrestrial companions formed in the system \citep{Ogihara2013}. Nevertheless, currently such terrestrial companions in the hot Jupiter systems have not discovered yet,
  this is because they may be under the detection limit.
  In the forthcoming, we hope that the hot Jupiter systems with low-mass companions could be discovered
  with the assistance of the improvement of observational precision and techniques.

  \item Furthermore, in the second sets, we also explore the 4:2:1 MMRs formation
  in the case of three terrestrial planets. Under type I migration, $\sim$ 17.1\% of all runs,
  indicate that terrestrial planets are evolved into 4:2:1 MMRs. However, as aforementioned, the probability that planets are trapped into MMRs depends on their initial separation and the speed of type I migration.
  Comparing the processes with and without giant planets, the
  timescales that three planet are in 4:2:1 MMRs in systems without
  giant planets are shorter than systems with giant planets. Additionally, eccentricities are exited to higher
  values in systems with giant planets than without giant planets.

\end{enumerate}

However, in the treatment of accretion scenario in MERCURY package \citep{Cham99},
the collisions between two bodies are assumed to be perfect
gravitational aggregations. \citet{lein12} pointed out that real collisions
could  suffer from partially accreting collision, hit-and-run impacts, or
graze-and-merge events, rather than a complete merger. Recently,
\citet{cham13} showed that fragmentation does have a
notable effect on accretion under consideration of fragmentation and
hit-and-run collisions \citep{lein12,gen12}.
The final stages of accretion are lengthened by the sweep up of collision fragments.
The planets that formed with fragmentation in the simulations have
smaller masses and lower eccentricities when compared to those
simulations without fragmentation. In this sense, the scenarios that include imperfect
merger collision will not alter the conclusions of this study.
In forthcoming work, we will take into account hit-and-run and fragmentation process
in our model for further investigation.

\section*{Acknowledgments}
We thank the anonymous referee for insightful comments
and good suggestions that helped to improve the contents.
This work is financially supported by National
Natural Science Foundation of China (Grants No. 11273068,11473073,11503092,11573073),
the Strategic Priority Research Program-The Emergence of
Cosmological Structures of the Chinese Academy of Sciences (Grant
No. XDB09000000), the innovative and interdisciplinary program by
CAS (Grant No. KJZD-EW-Z001), the Strategic Priority Research Program
on Space Science, CAS (Grant No. XDA04060901), the Natural Science Foundation of
Jiangsu Province (Grant No. BK20141509), and the Foundation of Minor
Planets of Purple Mountain Observatory.

\clearpage

%%Figure 1

\clearpage

%%Figure 2

\clearpage

%%Figure 3

\clearpage

%%Figure 4

\clearpage

%%Figure 5

%%Figure 6

\clearpage

%%Figure 7

   \clearpage

%%Figure 8

   \clearpage

%%Figure 9

\label{lastpage}


\begin{thebibliography}{99}

\bibitem[\protect\citeauthoryear{Adachi et al.}{1976}]{Ada76}Adachi,
I., Hayashi, C., \& Nakazawa, K., 1976, Prog. Theo. Phys. 56, 1756

\bibitem[\protect\citeauthoryear{Batalha et al.}{2013}]{batalha13}
Batalha, N.~M., et~al., 2013, ApJS, 204, 24

\bibitem[\protect\citeauthoryear{Batygin et al.}{2015}]{batygin15}
Batygin, K., Deck, K., \& Holman, M. J., 2015, AJ, 149, 167

\bibitem[\protect\citeauthoryear{Beaug\'e \& Nesvorn\'y}{2012}]{beau12}
Beaug\'e, C., \& Nesvorn\'y, D., 2012, ApJ, 751, 119

\bibitem[\protect\citeauthoryear{Bonfils et~al.}{2013}]{bon13}
Bonfils, X., et~al., 2013,  A\&A, 549, A109

\bibitem[\protect\citeauthoryear{Brunini \& Cionco}{2005}]{brun05}
Brunini, A., \& Cionco, R. G., 2005, Icarus, 177, 264

\bibitem[\protect\citeauthoryear{Cresswell \& Nelson}{2006}]{Cre06}
Cresswell, P., \& Nelson, R.~P., 2006,  A\&A, 450, 833

\bibitem[\protect\citeauthoryear{Chambers}{1999}]{Cham99}
Chambers, J. E., 1999, MNRAS, 304, 793

\bibitem[\protect\citeauthoryear{Chambers}{2001}]{cham01}
Chambers, J. E., 2001, Icarus 152, 205

\bibitem[\protect\citeauthoryear{Chambers}{2013}]{cham13}
Chambers, J. E., 2013, Icarus, 224, 43

\bibitem[\protect\citeauthoryear{Chapman et~al.}{2007}]{chapman07}
Chapman, C.~R., et~al., 2007, Icarus, 189, 233

%\bibitem[\protect\citeauthoryear{Chen et~al.}{2015}]{chen15}
%Chen, Y.~Y., Zhou J.~L., \& Ma Y.~H., 2015, Science China, 58, 3

\bibitem[\protect\citeauthoryear{Chiang \& Laughlin}{2013}]{chiang13}
Chiang, E., \& Laughlin, G., 2013, MNRAS, 431, 3444

\bibitem[\protect\citeauthoryear{Cohen et~al.}{2000}]{cohen00}
Cohen, B.~A., et~al., 2000, Science, 290, 1754

\bibitem[\protect\citeauthoryear{Cossou et~al.}{2013}]{cos13}
Cossou, C., et~al., 2013, A\&A, 553, L2

\bibitem[\protect\citeauthoryear{Cossou et~al.}{2014}]{cos14}
Cossou, C., et~al., 2014, A\&A, 569, A56

\bibitem[\protect\citeauthoryear{Dong \& Ji}{2013}]{dong13}
Dong, Y., \& Ji, J.~H., 2013, MNRAS, 430, 951

\bibitem[\protect\citeauthoryear{Fabrycky \& Tremaine}{2007}]{fab07}
Fabrycky, D. \& Tremaine, S., 2007, ApJ, 669, 1298


\bibitem[Fogg \& Nelson(2005)]{Fogg05} Fogg, M.~J., \& Nelson, R.~P.\ 2005, A\&A, 441, 791

\bibitem[Fogg \& Nelson(2009)]{Fogg09} Fogg, M.~J., \& Nelson, R.~P.\ 2009, A\&A, 498, 575


\bibitem[\protect\citeauthoryear{Ford \& Rasio}{2006}]{ford06}
Ford, E. B., \& Rasio, F. A., 2006, ApJL, 638, L45

\bibitem[\protect\citeauthoryear{Fressin et~al.}{2013}]{fres13}
Fressin, F., et~al., 2013, ApJ, 766, 81

\bibitem[\protect\citeauthoryear{Genda et al.}{2012}]{gen12}
Genda, H., Kokubo, E., \& Ida, S., 2012, ApJ, 744, 137

\bibitem[\protect\citeauthoryear{Goldreich \& Tremaine}{1979}]{GT79}
Goldreich, P., \& Tremaine, S., 1979, ApJ, 233, 857

\bibitem[\protect\citeauthoryear{Goldreich \& Tremaine}{1980}]{GT80}
Goldreich, P., \& Tremaine, S., 1980, ApJ, 241, 425

\bibitem[\protect\citeauthoryear{Haisch et al.}{2001}]{hai01}
Haisch, K.~E., Jr., Lada, E.~A., \& Lada, C.~J., 2001, ApJL, 553, L153

\bibitem[\protect\citeauthoryear{Hansen \& Murray}{2012}]{hansen12}
Hansen, B.~M.~S., \& Murray, N., 2012, ApJ, 751, 158

\bibitem[\protect\citeauthoryear{Hansen \& Murray}{2013}]{hansen13}
Hansen, B.~M.~S., \& Murray, N., 2013, ApJ, 775, 53

\bibitem[\protect\citeauthoryear{Hayashi}{1981}]{hay81}
Hayashi, C., 1981, Prog. THeor. Phys. Suppl., 70, 35

\bibitem[\protect\citeauthoryear{Howard et al.}{2010}]{how10}
Howard, A. W., et al., 2010, Science, 330, 653

\bibitem[\protect\citeauthoryear{Howard et al.}{2012}]{how12}
Howard, A.~W., et~al., 2012, ApJS, 201, 15

\bibitem[\protect\citeauthoryear{Ida \& Lin}{2004}]{idalin04}
Ida, S., \& Lin, D.~N.~C., 2004, ApJ, 604, 388

\bibitem[\protect\citeauthoryear{Ida \& Lin}{2008}]{IL08}
Ida, S., \& Lin, D.~N.~C., 2008, ApJ, 673, 487

\bibitem[\protect\citeauthoryear{Ida \& Makino}{1992}]{ida92}
Ida, S., \& Makino, J., 1992, Icarus, 96, 107

\bibitem[\protect\citeauthoryear{Izidoro et al.}{2014}]{izi14}
Izidoro, A., Morbidelli, A., \& Ramond, S. N., 2014, ApJ, 794, 11

\bibitem[\protect\citeauthoryear{Ji et al.}{2011}]{ji11}
Ji, J.~H., Jin, S., \& Tinney, C.~G., 2011, ApJL, 727, L5

\bibitem[\protect\citeauthoryear{Jin \& Ji}{2011}]{jin11}
Jin, S., \& Ji, J.~H., 2011, MNRAS, 418, 1335


\bibitem[\protect\citeauthoryear{Kokubo \& Ida}{1998}]{kok98}
Kokubo, E., \& Ida, S., 1998, Icarus, 131, 171

\bibitem[\protect\citeauthoryear{Kokubo \& Ida}{2002}]{Kok02}
Kokubo, E., \& Ida, S., 2002, ApJ, 581, 666


\bibitem[\protect\citeauthoryear{Lee \& Peale}{2002}]{lee02}
Lee M. H., Peale S. J., 2002, ApJ, 567, 596

\bibitem[\protect\citeauthoryear{Lee et al.}{2013}]{lee13}
Lee, M. H., Fabrycky, D., \& Lin, D. N. C., 2013, ApJL, 774, L52

%\bibitem[\protect\citeauthoryear{Lega et al.}{2013}]{lega13}
%Lega, E., Morbidelli, A., \& Nesvorn\'y, D., 2013, MNRAS, 431, 3494

\bibitem[\protect\citeauthoryear{Lega et al.}{2014}]{lega14}
Lega, E., et. al., 2014, MNRAS, 440, 683

\bibitem[\protect\citeauthoryear{Leinhardt \& Stewart}{2012}]{lein12}
Leinhardt, Z.M., \& Stewart, S.T., 2012, ApJ, 745,79


\bibitem[\protect\citeauthoryear{Lin \& Papaloizou}{1986}]{lp86}
Lin, D. N. C., \& Papaloizou, J. C. B., 1986, ApJ, 309, 846

\bibitem[\protect\citeauthoryear{Lin \& Papaloizou}{1993}]{lp93}
Lin, D. N. C., \& Papaloizou, J. C. B., 1993, in: E.H. Levy \& J.I.
Lunine (eds.), Protostars and Planets III, (Tucson: Unv. Arizona)

\bibitem[\protect\citeauthoryear{Lissauer et~al.}{2011}]{lis11}
Lissauer, J.~J., et~al., 2011, Nature, 470, 53

\bibitem[\protect\citeauthoryear{Lissauer et~al.}{2011b}]{lis11b}
Lissauer, J.~J., et~al., 2011, ApJS, 197, 8

\bibitem[\protect\citeauthoryear{Lovis et~al.}{2011}]{lov11}
Lovis, C., et~al., 2011,  A\&A, 528, A112

\bibitem[\protect\citeauthoryear{Mandell \& Sigurdsson}{2003}]{ms03}
Mandell, A. M., \& Sigurdsson, S., 2003, ApJ, 591, L111

\bibitem[\protect\citeauthoryear{Mandell et~al.}{2007}]{man07}
Mandell, A. M., Ramond, S. N., \& Sigurdsson, S., 2007, ApJ, 660, 823

\bibitem[\protect\citeauthoryear{Marcy et~al.}{2001}]{marcy01}
Marcy, G. W., et~al., 2001, ApJ, 556, 296

\bibitem[\protect\citeauthoryear{Marcy et~al.}{2014}]{marcy14}
Marcy, G. W., et~al., 2014, ApJS, 210, 20

\bibitem[\protect\citeauthoryear{Mart\'i et~al.}{2013}]{mar13}
Mart\'i, J. G., Giuppone, C. A., \& Beaug\'e, C., 2013, MNRAS, 433, 928

\bibitem[\protect\citeauthoryear{Mayor et~al.}{2011}]{may11}
Mayor, M., et~al., 2011, \emph{arXiv}:1109.2497

\bibitem[\protect\citeauthoryear{Morbidelli et~al.}{2007}]{morb07}
Morbidelli, A., et~al., 2007, AJ, 134, 1790

\bibitem[\protect\citeauthoryear{Morbidelli \& Crida}{2007b}]{morb07b}
Morbidelli, A., \& Crida, A., 2007, Icarus, 191, 158

\bibitem[\protect\citeauthoryear{Nelson et al.}{2016}]{nelson16}
Nelson, B.~E., et~al., 2016, MNRAS, 455, 2484

\bibitem[\protect\citeauthoryear{Ogihara \& Ida}{2009}]{Ogihara2009}
Ogihara, M., \& Ida, S. 2009, ApJ, 699, 824

\bibitem[\protect\citeauthoryear{Ogihara et al.}{2013}]{Ogihara2013}
Ogihara, M., Inutsuka, S. \& Kobayashi, H. 2013, ApJL, 778, 9

\bibitem[\protect\citeauthoryear{Ogihara et al.}{2015}]{Ogihara2015}
Ogihara, M., Morbidelli, A., \& Guillot, T. 2015, A\&A, 578, A36

\bibitem[\protect\citeauthoryear{Pierens \& Nelson}{2008}]{pn08}
Pierens, A., \& Nelson, R. P., 2008,  A\&A, 482, 333

%\bibitem[\protect\citeauthoryear{Papaloizou \& Terquem}{2010}]{pt10}
%$Papaloizou, J. C. B., \& Terquem, C., 2010, MNRAS, 405, 573

\bibitem[\protect\citeauthoryear{Raymond et al.}{2004}]{raym04}
Raymond  S. N., Quinn  T., Lunine  J. I., 2004, Icarus, 168, 1

\bibitem[\protect\citeauthoryear{Raymond et al.}{2006}]{raym06}
Raymond, S. N., et al., 2006, Science, 313, 1413

\bibitem[\protect\citeauthoryear{Raymond et al.}{2006b}]{raym06b}
Raymond, S. N., et al., 2006, Icarus, 183, 265

\bibitem[\protect\citeauthoryear{Raymond et al.}{2008}]{raym08}
Raymond, S. N., et al., 2008, MNRAS, 384, 663

\bibitem[\protect\citeauthoryear{Raymond et al.}{2008b}]{ray08b}
Raymond, S. N., et al., 2008, ApJ, 687, L107

\bibitem[\protect\citeauthoryear{Raymond \& Cossou}{2014}]{raymond14}
Raymond, S.~N., \& Cossou, C., 2014, MNRAS, 440, L11

\bibitem[\protect\citeauthoryear{Rivera et~al.}{2010}]{rivera10}
Rivera, E. J., et~al., 2010, ApJ, 719, 890

\bibitem[\protect\citeauthoryear{Rowe et al.}{2014}]{rowe14}
Rowe, J. F., et al., 2014, ApJ, 784, 45

\bibitem[\protect\citeauthoryear{Safronov}{1969}]{Saf69}
Safronov,V.S., 1969, Evolution of the Protoplanetary Cloud and
Formation of the Earth and the Planets (Moscow:Nauka)

\bibitem[\protect\citeauthoryear{Schneider et al.}{2011}]{Schn11}
Schneider J. et al., 2011,  A\&A, 532, A79

\bibitem[\protect\citeauthoryear{Steffen et al.}{2010}]{Ste10}
Steffen, J. H., Batalha, N. M., Borucki, W. J., et al., 2010, ApJ, 725, 1226

\bibitem[\protect\citeauthoryear{Tanaka \& Ida}{1999}]{Tan99}
Tanaka, H., \& Ida, S., 1999, Icarus, 139, 350

\bibitem[\protect\citeauthoryear{Tanaka et al.}{2002}]{Tan02}
Tanaka, H., Takeuchi, T., \& Ward, W.~R., 2002, ApJ, 565, 1257

\bibitem[\protect\citeauthoryear{Tera et~al.}{1974}]{tera74}
Tera, F., et~al., 1974, Earth and Planetary Science Letters, 22, 1

\bibitem[\protect\citeauthoryear{Terquem \& Papaloizou}{2007}]{terq07}
Terquem, C., \& Papaloizou, J. C. B., 2007, ApJ, 654, 1110


\bibitem[\protect\citeauthoryear{Tsiganis et~al.}{2005}]{tsiganis05}
Tsiganis, K., et~al., 2005, Nature, 435, 459

\bibitem[\protect\citeauthoryear{Udry et~al.}{2007}]{udry07}
Udry, S., et~al., 2007,  A\&A, 469, L43

\bibitem[\protect\citeauthoryear{Wang et al.}{2014}]{Wang14}
Wang, J., Xie, J.~W., Barclay, T., \& Fischer, D. A., 2014, ApJ, 783, 4

\bibitem[\protect\citeauthoryear{Wang et al.}{2012}]{wjz12}
Wang, S., Ji, J.~H., \& Zhou, J.~L., 2012, ApJ, 753, 170

\bibitem[\protect\citeauthoryear{Wang \& Ji}{2014}]{wang14}
Wang, S., \& Ji, J.~H., 2014, ApJ, 795, 85

\bibitem[\protect\citeauthoryear{Ward}{1997}]{Ward97}
Ward, W.~R., 1997, Icarus, 126, 261

\bibitem[\protect\citeauthoryear{Wetherill}{1980}]{wet80}
Wetherill, G. W., 1980, ARAA, 18, 77

\bibitem[\protect\citeauthoryear{Wetherill \& Stewart}{1989}]{weth89}
Wetherill, G. W., \& Stewart, G. R., 1989, Icarus, 77, 330

\bibitem[\protect\citeauthoryear{Zhang \& Zhou}{2010}]{zz10}
Zhang, H., \& Zhou, J. L., 2010, ApJ, 714, 532

\bibitem[\protect\citeauthoryear{Zhang \& Zhou}{2010b}]{zz10b}
Zhang, H., \& Zhou, J. L., 2010, ApJ, 719, 671

\bibitem[\protect\citeauthoryear{Zhang \& Ji}{2009}]{zha09}
Zhang, N., \& Ji, J.~H., 2009,  Science  in  China  Series G, 52(5), 794

\bibitem[\protect\citeauthoryear{Zhang et al.}{2010}]{zhang10}
Zhang, N., Ji, J.~H., \& Sun, Z., 2010, MNRAS, 405, 2016

\bibitem[\protect\citeauthoryear{Zhang et al.}{2014}]{zhang14}
Zhang, X.-J., Li, H., Li, S.-T., \& Lin, D.~N.~C., 2014, ApJL, 789, L23

\bibitem[\protect\citeauthoryear{Zhou et al.}{2005}]{zhou05}
Zhou, J.~L., Aarseth, S. J., et al., 2005, ApJ, 631, L85

\end{thebibliography}
\end{document}